\documentclass{elsart}

\usepackage{graphicx}

\begin{document}

\begin{frontmatter}
\title{Measurements of Six-Body Hadronic Decays of the $D^0$ Charmed Meson}

The FOCUS Collaboration

\author[ucd]{J.~M.~Link}
\author[ucd]{P.~M.~Yager}
\author[cbpf]{J.~C.~Anjos}
\author[cbpf]{I.~Bediaga}
\author[cbpf]{C.~G\"obel}
\author[cbpf]{A.~A.~Machado}
\author[cbpf]{J.~Magnin}
\author[cbpf]{A.~Massafferri}
\author[cbpf]{J.~M.~de~Miranda}
\author[cbpf]{I.~M.~Pepe}
% ...

\address[ucd]{University of California, Davis, CA 95616} 
\address[cbpf]{Centro Brasileiro de Pesquisas F\'\i sicas, Rio de Janeiro, RJ, Brasil} 
\address[cinv]{CINVESTAV, 07000 M\'exico City, DF, Mexico} 
\address[cu]{University of Colorado, Boulder, CO 80309} 
\address[fnal]{Fermi National Accelerator Laboratory, Batavia, IL 60510} 
\address[fras]{Laboratori Nazionali di Frascati dell'INFN, Frascati, Italy I-00044}
\address[guan]{University of Guanajuato, 37150 Leon, Guanajuato, Mexico} 
\address[ui]{University of Illinois, Urbana-Champaign, IL 61801} 
\address[iu]{Indiana University, Bloomington, IN 47405} 
\address[korea]{Korea University, Seoul, Korea 136-701}
\address[korea2]{Kyungpook National University, Taegu, Korea 702-701}
\address[milan]{INFN and University of Milano, Milano, Italy} 
\address[nc]{University of North Carolina, Asheville, NC 28804} 
\address[pavia]{Dipartimento di Fisica Nucleare e Teorica and INFN, Pavia, Italy} 
\address[pr]{University of Puerto Rico, Mayaguez, PR 00681} 
\address[sc]{University of South Carolina, Columbia, SC 29208} 
\address[ut]{University of Tennessee, Knoxville, TN 37996} 
\address[vu]{Vanderbilt University, Nashville, TN 37235} 
\address[wisc]{University of Wisconsin, Madison, WI 53706}

\endnote{\small See http://www-focus.fnal.gov/authors.html for
additional author information}

\begin{abstract}
Using data collected by the FOCUS experiment at Fermilab, we report the discovery of the decay modes $D^0 \rightarrow K^- \pi^+ \pi^+ \pi^+ \pi^- \pi^-$ and $D^0 \rightarrow \pi^+ \pi^+ \pi^+ \pi^- \pi^- \pi^-$.  With a sample of $48 \pm 10$ reconstructed $D^0 \rightarrow K^- \pi^+ \pi^+ \pi^+ \pi^- \pi^-$ decays and $149 \pm 17$ reconstructed $D^0 \rightarrow \pi^+ \pi^+ \pi^+ \pi^- \pi^- \pi^-$ decays, we measure the following relative branching ratios: 

\noindent ${\Gamma (D^0 \rightarrow K^- \pi^+ \pi^+ \pi^+ \pi^- \pi^- ) / \Gamma (D^0 \rightarrow K^- \pi^+ \pi^+ \pi^- )} = (2.70 \pm 0.58 \pm 0.38) \times 10^{-3}$

\noindent ${\Gamma (D^0 \rightarrow \pi^+ \pi^+ \pi^+ \pi^- \pi^- \pi^- ) / \Gamma (D^0 \rightarrow K^- \pi^+ \pi^+ \pi^- )} = (5.23 \pm 0.59 \pm 1.35) \times 10^{-3}$

\noindent ${\Gamma (D^0 \rightarrow \pi^+ \pi^+ \pi^+ \pi^- \pi^- \pi^- ) / \Gamma (D^0 \rightarrow K^- \pi^+ \pi^+ \pi^+ \pi^- \pi^- )} = 1.93 \pm 0.47 \pm 0.48$

\noindent The first errors are statistical and the second are systematic.  The branching fraction of the Cabibbo suppressed six-body decay mode is measured to be a factor of two higher than the branching fraction of the Cabibbo favored six-body decay mode. 
\end{abstract}
\end{frontmatter}

\section{Introduction}

Hadronic decays of charmed mesons have been extensively studied in recent years.  However, six-body hadronic decays of the $D^0$ have not been previously observed; only an upper limit exists for the $D^0 \rightarrow \pi^+ \pi^+ \pi^+ \pi^- \pi^- \pi^-$ branching fraction~\cite{Barlag:1992ww}.  In this paper, we present the first branching ratio measurements of the $D^0 \rightarrow K^- \pi^+ \pi^+ \pi^+ \pi^- \pi^-$ and $D^0 \rightarrow \pi^+ \pi^+ \pi^+ \pi^- \pi^- \pi^-$ decay modes.  Charge conjugate states are implicitly included and we use the abbreviations $D^0 \rightarrow K5\pi$, $D^0 \rightarrow 6\pi$, and $D^0 \rightarrow K3\pi$ for the fully charged states.

The fixed-target charm photoproduction experiment FOCUS collected data during the 1996--1997 fixed-target run at Fermilab.  The FOCUS detector is a large aperture spectrometer with excellent vertexing and particle identification capabilities.  A photon beam is derived from the bremsstrahlung of secondary electrons and positrons produced from the 800 GeV/$c$ Tevatron proton beam.  The photon beam interacts with a segmented beryllium-oxide target. The average photon energy for the interactions collected for the measurements we report is 180 GeV.  Charged particles are tracked by two systems of silicon microvertex detectors.  The upstream system~\cite{Link:2002zg}, consisting of 4 planes (two views in two stations),  is interleaved with the experimental target, while the other system lies downstream of the target and consists of twelve planes of microstrips arranged in four stations of three views.  These detectors provide high resolution separation of production and decay vertices.  The momentum of a charged particle is determined by measuring its deflections in two analysis magnets of opposite polarity with five stations of multiwire proportional chambers.  Three multicell threshold \v Cerenkov counters are used to discriminate between electrons, pions, kaons, and protons.

\section{Signals and selection criteria}

A candidate driven vertexing algorithm~\cite{Frabetti:1992au} is used to reconstruct $D^0$ decays into six-body final
states.  A $D^0$ candidate consists of six tracks in an event that have zero total charge and form a vertex with at least a 2\% confidence level.  The momentum vector of the $D^0$ candidate, formed from the momenta of the six tracks, is then
intersected with at least one other track in the event to form the $D^0$ production vertex; the confidence level of this vertex
is required to be at least 1\%.  Additional cuts are applied based on event geometry and particle identification.  To minimize systematic errors, identical cuts are used on the two six-body decay modes and on the normalizing mode, except that the $D^0 \rightarrow 6\pi$ mode has no kaon identification cut.  Our most effective cut for reducing non-charm backgrounds is a significance of detachment cut that requires the separation, $\ell$, between the $D^0$ production and decay vertices divided by its error, $\sigma_{\ell}$, to be greater than some threshold, in our case $\ell/\sigma_{\ell} > 13$. The $D^0$ decay vertex is also required to be located outside of material in the target region by at least 2 standard deviations, which serves to reduce backgrounds from secondary interactions.  \v{C}erenkov particle identification is done using a $\chi^2$-like variable $W_i = -2 \ln \hbox{Likelihood}(i)$, where $i$ ranges over electron, pion, kaon, and proton hypotheses~\cite{Link:2001pg}.  For each pion candidate, we require $\min\{W_e,W_K,W_p\} - W_{\pi} > -4$, which requires that each pion candidate is not highly favored to be an electron, kaon, or proton rather than a pion.  For the kaon candidate in $D^0 \rightarrow K5\pi$ and in the normalizing mode, we require that the kaon hypothesis is more likely than the pion hypothesis with the cut $W_{\pi} - W_K > 3$.  Finally, the largest confidence level that one of the tracks from the decay vertex intersects the production vertex is required to be less than 25\%.

The invariant mass distributions of the $D^0$ candidates that satisfy these criteria are plotted in Figure 1.
\begin{figure}
\centering
\includegraphics[width=5.5in,bb=0 274 540 524]{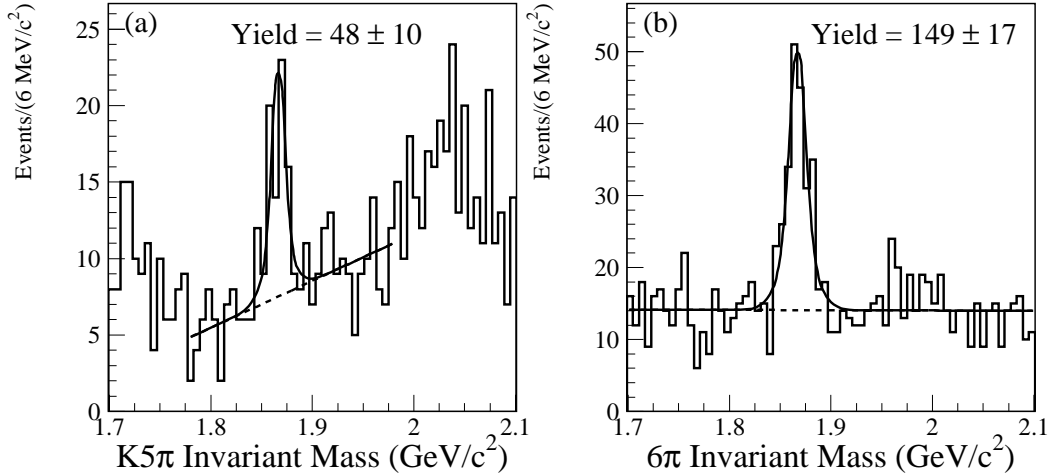}
\caption {Invariant mass distributions of (a) $K5\pi$ and (b) $6\pi$.}
\end{figure}
The $D^0 \rightarrow K5\pi$ mass plot is fit with a linear polynomial plus two Gaussians with the same mean but different widths.  We fit with two Gaussians because the mass resolution varies with momentum and the position of the decay vertex; the sum of two Gaussians provides a much better approximation to this situation than a single Gaussian.  The widths of the two Gaussians and their relative yields are fixed to values obtained from a Monte Carlo simulation (61\% of the total yield is in a Gaussian shape with $\sigma = 5.9 \hbox{ MeV}/c^2$ and 39\% of the total yield is in a Gaussian shape with $\sigma = 13.1 \hbox{ MeV}/c^2$).  The fit returns a signal yield of $48 \pm 10$ events.  Based on studies of reflections above and below the signal, we choose to fit over the range 1.78 GeV/$c^2$ to 1.98 GeV/$c^2$.  The reflection below 1.78 GeV/$c^2$ is consistent with partial reconstruction of seven-body final states from $D^+ \rightarrow K6\pi$ and from the decay chain $D^0 \rightarrow K^- \pi^+ \eta'$, $\eta' \rightarrow \pi^+ \pi^- \eta$, $\eta \rightarrow \pi^+ \pi^- \pi^0$ or $\eta \rightarrow \pi^+ \pi^- \gamma$, which yields the same final state as $K5\pi$ with an additional $\pi^0$ or $\gamma$.  The structure above 2.0 GeV/$c^2$ is due to $D^+ \rightarrow K^- \pi^+ \pi^+ \pi^+ \pi^-$ with a random $\pi^-$ intersecting the decay vertex.

The $D^0 \rightarrow 6\pi$ mass plot is also fit with a linear polynomial plus two Gaussians with the same mean whose widths and relative yield are fixed to values from a Monte Carlo simulation (60\% of the total yield is in a Gaussian shape with $\sigma = 7.6 \hbox{ MeV}/c^2$ and 40\% of the total yield is in a Gaussian shape with $\sigma = 16.2 \hbox{ MeV}/c^2$).  The fit returns $149 \pm 17$ events.

We measure the branching ratios of $D^0 \rightarrow K^- \pi^+ \pi^+ \pi^+ \pi^- \pi^-$ and $D^0 \rightarrow \pi^+ \pi^+ \pi^+ \pi^- \pi^- \pi^-$ relative to the high statistics mode $D^0 \rightarrow K^- \pi^+ \pi^+ \pi^-$.  The $D^0 \rightarrow K3\pi$ normalizing mode is fit in the same way as the two six-body modes, and the fit returns $70\,466 \pm 277$ signal events.  We also directly measure the relative branching ratio of the two six-body decay modes $\Gamma(D^0 \rightarrow \pi^+ \pi^+ \pi^+ \pi^- \pi^- \pi^-)/\Gamma(D^0 \rightarrow K^- \pi^+ \pi^+ \pi^+ \pi^- \pi^-)$ in order to take into account any correlations in systematic errors on the two modes.  From Monte Carlo simulations using a nonresonant model of each six-body decay mode, we compute the relative efficiencies ${\epsilon(D^0 \rightarrow K5\pi) \over \epsilon(D^0 \rightarrow K3\pi)} = 0.254 \pm 0.004$, ${\epsilon(D^0 \rightarrow 6\pi) \over \epsilon(D^0 \rightarrow K3\pi)} = 0.405 \pm 0.004$, and ${\epsilon(D^0 \rightarrow 6\pi) \over \epsilon(D^0 \rightarrow K5\pi)} = 1.596 \pm 0.027$.  The resulting branching ratio measurements are shown in Table 1.
\begin{table}
\centering
\caption {Branching ratio measurements.  The first error is statistical and the second is systematic.}
\begin{tabular}{lr} \hline
Decay Mode & Branching Ratio \\ \hline
${\Gamma (D^0 \rightarrow K^- \pi^+ \pi^+ \pi^+ \pi^- \pi^- ) \over \Gamma (D^0 \rightarrow K^- \pi^+ \pi^+ \pi^- )}$        & $(2.70 \pm 0.58 \pm 0.38) \times 10^{-3}$ \\
${\Gamma (D^0 \rightarrow \pi^+ \pi^+ \pi^+ \pi^- \pi^- \pi^- ) \over \Gamma (D^0 \rightarrow K^- \pi^+ \pi^+ \pi^- )}$        & $(5.23 \pm 0.59 \pm 1.35) \times 10^{-3}$ \\
${\Gamma (D^0 \rightarrow \pi^+ \pi^+ \pi^+ \pi^- \pi^- \pi^- ) \over \Gamma (D^0 \rightarrow K^- \pi^+ \pi^+ \pi^+ \pi^- \pi^- )}$      & $1.93 \pm 0.47 \pm 0.48$\\ \hline
\end{tabular}
\end{table}
Since the $D^0 \rightarrow 6\pi$ mode is Cabibbo suppressed while the $D^0 \rightarrow K5\pi$ mode is Cabibbo favored, one might expect the relative branching ratio of these two modes to be about $\tan^2 \theta_C \approx 0.05$, where $\theta_C$ is the Cabibbo angle.  Our measurement of their relative branching ratio is ${\Gamma (D^0 \rightarrow 6\pi) \over \Gamma (D^0 \rightarrow K 5\pi)} = 1.93 \pm 0.47 \hbox{ (stat.)} \pm 0.48 \hbox{ (sys.)}$.

\bigskip

\section{Systematic errors}

To check for bias in our selection of cuts, we have studied the sensitivity of the results to cut selection by individually varying each cut over a reasonable range of values.  The branching ratio measurements are stable as the cuts are varied.  We have investigated a number of sources of systematic uncertainty in the branching ratio measurements.  These sources are described below, and the systematic errors associated with them are listed in Table 2.
\begin{table}
\centering
\caption{Systematic error contributions as a percentage of the branching ratio.}
\begin{tabular}{lccc} \hline
Source & ${\Gamma (D^0 \rightarrow K5\pi) \over \Gamma (D^0 \rightarrow K3\pi)}$ & ${\Gamma (D^0 \rightarrow 6\pi) \over \Gamma (D^0 \rightarrow K3\pi)}$ & ${\Gamma (D^0 \rightarrow 6\pi) \over \Gamma (D^0 \rightarrow K5\pi)}$ \\ \hline
Run period split & 13.0\% & 25.3\% & 24.2\% \\
Momentum split & 0.0\% & 0.0\% & 0.0\% \\
Fitting & 3.4\% & 4.1\% & 5.3\% \\
Subresonances & 2.8\% & 2.5\% & 3.6\% \\
6-body vs. 4-body & 2.8\% & 2.8\% & -- \\ \hline
Total & 14.0\% & 25.9\% & 25.0\% \\ \hline
\end{tabular}
\end{table}

We quantify the systematic uncertainty on the efficiency due to Monte Carlo simulation using the split sample procedure based on the $S$-factor method employed by the Particle Data Group~\cite{Link:2002hi,PDBook}.  The data set is split into independent subsamples by reconstructed $D^0$ momentum and by early and late runs, which have different target and silicon microvertex detector configurations.  Because of our limited statistics, the splits are done one variable at a time.  We measure the branching ratio for each independent subsample and assess whether the subsample measurements are consistent with a single value by examining the $\chi^2$.  If $\chi^2/(\hbox{degrees of freedom}) > 1$, we scale up the errors such that $\chi^2/\hbox{d.o.f.} = 1$.  If the scaled error on the weighted average of the subsample measurements exceeds the statistical error on the whole sample measurement, we define the split sample systematic error to be the difference in quadrature between the scaled error and the statistical error.

We have studied the dependence of the results on the fitting procedure by fitting the histograms in several different ways:  using one Gaussian instead of two, different bin sizes, and different background parameterizations.  In the $D^0 \rightarrow K5\pi$ case, we also used fit functions that included reflection shapes obtained from Monte Carlo simulations.  The sample standard deviation of the branching ratio measurements from the different fit variants is taken as the fit variant contribution to the systematic error.

Because the resonance substructures of $D^0 \rightarrow K5\pi$ and $D^0 \rightarrow 6\pi$ are unknown and the subresonance model used in the Monte Carlo affects the reconstruction efficiency, we compute the efficiencies for several subresonance models and use the sample standard deviation of the resulting branching ratios as a contribution to the systematic error.  The subresonance models used for $D^0 \rightarrow K5\pi$ are $D^0 \rightarrow K^- a_1(1260)^+$, $D^0 \rightarrow \overline{K}^{*0} \rho(1450)$ ($\rho(1450) \rightarrow 4\pi$), and $D^0 \rightarrow \overline{K}^{*0} 4\pi $ in addition to a nonresonant model.  For $D^0 \rightarrow 6\pi$, the subresonance model $D^0 \rightarrow \pi^- a_1(1260)^+$ is used in addition to a nonresonant model.  For subresonance models of both modes involving the $a_1(1260)^+$, we compute the efficiency for each of three different models for the decay of the $a_1(1260)^+$:  $a_1(1260)^+ \rightarrow f_2(1270) \pi^+$, $a_1(1260)^+ \rightarrow f_0(1370) \pi^+$, and $a_1(1260)^+ \rightarrow \rho(1450) \pi^+$, where the $f_2(1270)$, $f_0(1370)$, and $\rho(1450)$ decay to four charged pions.  The mass and width of the $a_1(1260)^+$ are assumed to be 1230 MeV/$c^2$ and 400 MeV/$c^2$, respectively.

We also include a systematic error contribution from differences in absolute tracking efficiencies for six-body versus four-body final states.  The total systematic error is obtained by adding the different contributions in quadrature.

\section{Conclusion}

We have presented the first measurements of six-body hadronic decays of the $D^0$.  The results are:

${\Gamma (D^0 \rightarrow K^- \pi^+ \pi^+ \pi^+ \pi^- \pi^- ) \over \Gamma (D^0 \rightarrow K^- \pi^+ \pi^+ \pi^- )} = (2.70 \pm 0.58 \pm 0.38) \times 10^{-3}$

${\Gamma (D^0 \rightarrow \pi^+ \pi^+ \pi^+ \pi^- \pi^- \pi^- ) \over \Gamma (D^0 \rightarrow K^- \pi^+ \pi^+ \pi^- )} = (5.23 \pm 0.59 \pm 1.35) \times 10^{-3}$

${\Gamma (D^0 \rightarrow \pi^+ \pi^+ \pi^+ \pi^- \pi^- \pi^- ) \over \Gamma (D^0 \rightarrow K^- \pi^+ \pi^+ \pi^+ \pi^- \pi^- )} = 1.93 \pm 0.47 \pm 0.48$

The relative branching ratio of the two six-body decay modes is much higher than one might expect from Cabibbo suppression.  Theoretical discussion of many-body charm decays has suggested a ``vector-dominance model'' in which a charmed meson emits a $W^{\pm}$ which hadronizes into a charged vector, axial-vector, or pseudoscalar meson~\cite{Lipkin:2000gz}.  Studies of five-body charm decays by FOCUS have provided evidence for this model with five-body decays of the $D^0$, $D^+$, and $D_s^+$ being dominated by quasi-two-body decays involving the $a_1(1260)^{\pm}$~\cite{Link:2002mm,Link:2003rj}.  Our result for $\Gamma (D^0 \rightarrow 6\pi) / \Gamma (D^0 \rightarrow K5\pi)$ may be qualitatively explained by the hypothesis that six-body decays of the $D^0$ proceed primarily through quasi-two-body decays involving an $a_1(1260)^+$.  The decay channels of the $a_1(1260)^+$ that can result in five charged pions are $f_2(1270) \pi^+$, $f_0(1370) \pi^+$, and $\rho(1450) \pi^+$.  If the $D^0$ decays to $K^- a_1(1260)^+$, then only fractions of the widths of the $f_2(1270)$, $f_0(1370)$, and  $\rho(1450)$ are available for the decay of the $a_1(1260)^+$, resulting in a significant suppression of six-body final states involving a kaon compared with six pion final states from the decay $D^0 \rightarrow \pi^- a_1(1260)^+$.

We wish to acknowledge the assistance of the staffs of Fermi National
Accelerator Laboratory, the INFN of Italy, and the physics departments of the
collaborating institutions. This research was supported in part by the U.~S.
National Science Foundation, the U.~S. Department of Energy, the Italian
Istituto Nazionale di Fisica Nucleare and Ministero della Istruzione
Universit\`a e Ricerca, the Brazilian Conselho Nacional de Desenvolvimento
Cient\'{\i}fico e Tecnol\'ogico, CONACyT-M\'exico, and the Korea Research
Foundation of the Korean Ministry of Education.

\bibliographystyle{elsart-num}
\bibliography{sixbody_plb}

\end{document}